\documentclass[twocolumn,showpacs,preprintnumbers,amsmath,amssymb]{revtex4}
\usepackage{tabularx,graphicx}\begin{document}
\newcommand{\beq}{\begin{equation}}
\newcommand{\eeq}{\end{equation}}
\newcommand{\beqn}{\begin{eqnarray}}
\newcommand{\eeqn}{\end{eqnarray}}
\newcommand{\bmath}{\begin{subequations}}
\newcommand{\emath}{\end{subequations}}
\title{Charge expulsion, Spin Meissner effect, and charge inhomogeneity in  superconductors}
\author{J. E. Hirsch }
\address{Department of Physics, University of California, San Diego\\
La Jolla, CA 92093-0319}

\begin{abstract} 
Superconductivity occurs in systems that have a lot of negative charge: the highly negatively charged 
$(CuO_2)^=$ planes in the cuprates, negatively charged $(FeAs)^-$ planes in the iron arsenides, and negatively charged $B^-$  planes in magnesium diboride. And, in the nearly filled (with negative electrons) bands of almost all superconductors, as evidenced by their positive Hall coefficient in the normal state. No explanation for this charge asymmetry is provided by the conventional theory of superconductivity, within which the $sign$ of electric charge plays no role.  Instead, the sign of the charge carriers plays a key role in the theory of hole superconductivity, according to which metals become superconducting because they are driven to expel negative charge (electrons) from their interior. This is why NIS tunneling spectra are asymmetric, with larger current for negatively biased samples. The theory also offers a  compelling explanation of the Meissner effect: as electrons are expelled towards the surface in the presence of a magnetic field, the Lorentz force imparts them with azimuthal velocity, thus generating the surface Meissner current that screens the interior magnetic field. In type II superconductors, the Lorentz force acting on expelled electrons that don't reach the surface gives rise to the azimuthal velocity of the vortex currents. In the absence of applied magnetic field, expelled electrons still acquire azimuthal velocity, due to the spin-orbit interaction, in opposite direction for spin-up and spin-down electrons: the "Spin Meissner effect". This results in a macroscopic spin current flowing near the surface of superconductors in the absence of applied fields, of magnitude equal to the critical charge current
(in appropriate units). Charge expulsion also gives rise to an interior outward-pointing electric field and to excess negative charge near the surface. In strongly type II superconductors this physics should give rise to charge inhomogeneity and spin currents throughout the interior of the superconductor, to large sensitivity to (non-magnetic) disorder and to a strong tendency to phase separation.

   \end{abstract}
\maketitle 
\section{introduction}

In a perfectly conducting fluid, magnetic field lines move with the fluid (Alfven's theorem\cite{alfven}). Thus it is natural to infer that the expulsion of magnetic field lines from the
interior of a metal making a transition to the superconducting state (Meissner
effect\cite{meissner})
is likely to be  associated with {\it radially outward motion of electric charge}\cite{lorentz}. 
However, the conventional theory of superconductivity predicts no  radial motion of charge in the transition to  superconductivity\cite{tinkham}.
Rather, BCS-London theory postulates that a spontaneous $azimuthal$ charge motion near the surface is generated
(Meissner current)  to compensate the
  magnetic field in the interior, without however explaining what is the
driving force for such  a motion nor how angular momentum is conserved\cite{lenz}.

Instead, the theory of hole superconductivity\cite{hole} predicts
that superconductors expel negative charge from their interior towards the surface as they enter the superconducting state\cite{exp,undr} to lower their kinetic energy\cite{hole0}
associated with quantum confinement\cite{holem1},
whether or not an external magnetic field is present. In the
presence of an external magnetic field, the radial charge motion will 'drag' the magnetic field lines with it as in a classical plasma\cite{alfven}: the magnetic Lorentz force on a radially
moving charge acts in the azimuthal direction, and the deflected electron
motion generates a magnetic field in direction opposite to the applied one. Thus, the theory offers a
'dynamical' explanation of the
Meissner effect\cite{missing}.    In the absence of applied magnetic field, the radial outflow of charge gives rise to a spontaneous spin current\cite{sc}, predicted to
exist in the ground state of all superconductors\cite{sm}.

The hypothesis that the transition to superconductivity is associated with expulsion of negative charge is supported by the observation that 
high temperature superconductivity appears to be favored   in materials with substructures that have $excess$ negative charge, namely the
$(CuO_2)^=$ planes in the cuprates,  the $(FeAs)^-$ planes in the iron arsenides, and the  $B^-$  planes in magnesium diboride. 
The observation that in 
high $T_c$ materials normal-insulator-superconductor tunneling spectra are asymmetric, with larger current for a $negatively$ biased sample\cite{tunnasymexp,hole7},
is further evidence that  superconductors have a tendency to expel negative charge.
Finally, superconducting materials almost always exhibit a $positive$ Hall coefficient in the normal state\cite{hall}, which indicates
electronic bands almost full with negative electrons.

Superconductivity arises in our theory   when the Fermi level is close to the top of a band\cite{hole1},
and $T_c$ is enhanced when the ions are negatively charged\cite{hole2}. 
Electron-hole asymmetric electronic polaron models describe the physics of pair formation\cite{hole3}, which in the low energy sector reduce to a Hubbard model with correlated
hopping\cite{hole4}. The pair (bipolaron) is lighter than the single polaron in these models\cite{hole5} because the hopping amplitude increases with increasing local
hole occupation due to electron-hole asymmetry\cite{hole6},
and this effect promotes pairing of hole carriers\cite{hole4}. 
The models are derived from basic atomic physics considerations of wide generality\cite{hole6}, and the theory is proposed to apply to all superconducting materials\cite{all}.

\section{the two routes to the Meissner effect}

The fact that the Meissner effect is unexplained by the conventional theory is not generally recognized\cite{missing}. The canonical momentum of an electron with 
superfluid velocity $\vec{v}_s$  is
\beq
\vec{p}=m_e\vec{v}_s+\frac{e}{c}\vec{A}
\eeq
with $\vec{A}$ the magnetic vector potential. In the BCS ground state the expectation value $<\vec{p}>=0$, hence the superfluid velocity is given by
\beq
\vec{v}_s=-\frac{e}{m_e c}\vec{A}=-\frac{e\lambda_L}{m_ec}\vec{B}\times\hat{n}
\eeq
The second equality in Eq. (2) applies to a cylindrical geometry, where $\hat{n}$ is the outward pointing normal of the lateral surface of the cylinder and $\vec{B}$ is the magnetic field along the axis of the cylinder. The London penetration depth $\lambda_L$ is given by\cite{tinkham}
\beq
\frac{1}{\lambda_L^2}=\frac{4\pi n_s e^2}{m_e c^2}
\eeq
where $n_s$ is the superfluid density.

Eq. (2) embodies the Meissner effect\cite{tinkham}. However the BCS 'explanation' just outlined does not explain how the electrons are driven to acquire this 
velocity starting from a normal state where the average velocity is zero in the presence of a static magnetic field, nor how the mechanical angular momentum of the
carriers of the Meissner current is compensated\cite{missing}.

\subsection{Meissner current from orbit expansion}
Consider an electron that moves radially outward from the axis of a cylinder in the presence of a  magnetic field $\vec{B}$ parallel to the cylinder. The equation of motion is
\beq
m_e\frac{d\vec{v}}{dt}=\frac{e}{c}\vec{v}\times\vec{B}+\vec{F}_r
\eeq
where the first term is the magnetic Lorentz force and the second term is a radial force arising from ``quantum pressure'' that drives the electron outward\cite{holem1}.
 From Eq. (4),
\beq
\vec{r}\times\frac{d\vec{v}}{dt}=\frac{e}{m_ec}\vec{r}\times (\vec{v}\times\vec{B})
\eeq
where $\vec{r}$ is in the plane perpendicular to the axis of the cylinder. Hence $\vec{r}\cdot\vec{B}=0$ and 
$\vec{r}\times (\vec{v}\times\vec{B})=-(\vec{r}\cdot\vec{v})\vec{B}$, and
\beq
\frac{d}{dt}(\vec{r}\times\vec{v})=-\frac{e}{m_ec}(\vec{r}\cdot\vec{v})\vec{B}=-\frac{e}{2m_ec}(\frac{d}{dt}r^2)\vec{B}
\eeq
so that $\vec{r}\times\vec{v}=-(e/2m_e c)r^2\vec{B}$, and the acquired azimuthal velocity in moving out a distance $r$ is
\beq
v_\phi=-\frac{e}{2m_ec}rB
\eeq
Thus, to acquire the azimuthal speed Eq. (2) needed for the Meissner current requires the action of the Lorentz force {\it over a radially outgoing motion
 to   radius $r=2\lambda_L$}. 
 
\subsection{Meissner current from Faraday induction}
For an electron orbiting in a circular orbit of radius $r$, as an external magnetic field perpendicular to the orbit is applied, an azimuthal electric field
$E=(r/2c)\partial B/\partial t$ is generated by Faraday's law, and the velocity of the electron changes as
\beq
\frac{dv}{dt}=\frac{eE}{m_e}=\frac{er}{2m_ec}\frac{\partial B}{\partial t}
\eeq
so that for a magnetic field increasing from $0$ to $B$ the extra velocity acquired is
\beq
\Delta v=\frac{er}{2m_ec}B
\eeq
which reduces to Eq. (2) if and only if the orbit has radius $r=2\lambda_L$.
Hence, the hallmark property of superconductors, that the same Meissner current Eq. (2) results when a magnetic field is applied to an already superconducting metal or when a normal metal becomes superconducting in a pre-existent magnetic field, can be understood from the assumption that superconducting electrons reside in mesoscopic orbits of radius $2\lambda_L$. A parallel reasoning leads to the development of a ground state {\it spin current} in the absence of
applied fields\cite{sm} (Spin Meissner effect) as we discuss in the following section.

\section{the two routes to the Spin Meissner effect}
The superconducting condensate carries a charge density $en_s$. Thus, in a charge-neutral system the superfluid moves in a compensating background 
of positive charge density $\rho=|e|n_s$. The interaction of the moving magnetic moments of the electrons with the positive background leads to a universal spin Meissner current 
with speed of magnitude\cite{sm}
\beq
v_\sigma^0=\frac{\hbar}{4m_e\lambda_L}
\eeq
as we will shows in what follows, which parallels the discussion in the previous section.

\subsection{Spin Meissner current from orbit expansion}
Consider a magnetic moment $\vec{\mu}$ along the $z$ direction that moves radially outward with velocity $\vec{v}$. It is equivalent to an electric
dipole moment\cite{dipole}
\beq
\vec{p}=\frac{\vec{v}}{c}\times\vec{\mu}
\eeq
In the presence of the radial electric field of the cylinder
\beq
\vec{E}=2\pi \rho \vec{r}=2\pi |e|n_s\vec{r}
\eeq
the electric dipole experiences a torque
\beq
\vec{\tau}=\vec{p}\times\vec{E}=(\frac{\vec{v}}{c}\times\vec{\mu})\times\vec{E}=-2\pi |e|n_s\vec{r}\times(\frac{\vec{v}}{c}\times\vec{\mu})
\eeq
which causes a change in its angular momentum
\beq
\frac{d\vec{L}}{dt}=m_e\frac{d}{dt}(\vec{r}\times\vec{v})=\vec{\tau}
\eeq
Hence
\beq
\vec{r}\times\frac{d\vec{v}}{dt}=\frac{2\pi en_s}{m_e}\vec{r}\times(\frac{\vec{v}}{c}\times\vec{\mu})
\eeq
Eq. (15) is identical to Eq. (5) if we define the 'effective' magnetic field
\beq
\vec{B}_\sigma=2\pi n_s\vec{\mu}
\eeq
and hence leads to the azimuthal velocity Eq. (7) with $B_\sigma$ replacing $B$
\beq
v_\phi=-\frac{\pi e n_s}{m_e c}r\mu_B
\eeq
with $\mu_B=|e|\hbar/2m_e c$ the Bohr magneton, so that
\beq
v_\phi=\frac{\pi n_s e^2\hbar r}{2m_e^2c^2}=\frac{\hbar r}{8m_e \lambda_L^2}
\eeq
where we have used Eq. (3) for the second equality in Eq. (18). The two electrons in a Cooper pair 
have opposite spin and orbit in opposite directions.
The orbital angular momentum of each electron  is
\beq
l=m_e r v_\phi=\frac{\hbar r^2}{8 \lambda_L^2}
\eeq
For $r=2\lambda_L$, the azimuthal velocity Eq. (18) reduces to   Eq. (10) and the orbital angular momentum is
\beq
l=\frac{\hbar}{2}  .
\eeq
For any other value of $r$, the orbital angular momentum Eq. (19) is $not$ $\hbar/2$. 

\subsection{Spin Meissner current from Maxwell induction}
The same result for the azimuthal velocity Eq. (10) is obtained through a reasoning paralleling the second route to the Meissner
current (Sect. IIB).

Consider a magnetic moment $\vec{\mu}$ in an orbit of radius $r$, with $\vec{\mu}$ oriented perpendicular to the plane of the orbit. Assume a 
radial electric field grows from  $0$  to a final value $\vec{E}$. According to Ampere-Maxwell's law a magnetic field is induced by the
varying electric field, satisfying
\beq
\vec{\nabla}\times\vec{B}=\frac{1}{c}\frac{\partial \vec{E}}{\partial t}
\eeq
 which exerts an azimuthal force on the magnetic moment
 \beq
 \vec{F}=m_e\frac{d\vec{v}}{dt}=\vec{\nabla}(\vec{\mu}\cdot\vec{B})
 \eeq
 We have
 \beq
 \vec{\nabla}(\vec{\mu}\cdot\vec{B})=
 (\vec{\mu}\cdot\vec{\nabla})\vec{B}+\vec{\mu}\times(\vec{\nabla}\times\vec{B})
 \eeq
 In the geometry under consideration the first term in Eq. (23) is half the second term and points in opposite direction, so that
  \beq
 \vec{F}=m_e\frac{d\vec{v}}{dt}=\frac{1}{2}\vec{\mu}\times(\vec{\nabla}\times\vec{B})=\frac{1}{2c}\frac{\partial}{\partial t}(\vec{\mu}\times\vec{E})
 \eeq
 and the azimuthal velocity acquired is
 \beq
 \vec{v}_\phi=\frac{1}{2m_e c}\vec{\mu}\times\vec{E}
 \eeq
 and for the electric field given by Eq. (12)
 \beq
 \vec{v}_\phi=\frac{\pi |e|n_s}{m_e c}\vec{\mu}\times\vec{r}
 \eeq
 or
 \beq
 v_\phi=\frac{\pi n_s e^2\hbar r}{2m_e^2c^2}=\frac{\hbar r}{8m_e \lambda_L^2}
\eeq
in agreement with Eq. (18). Thus, just like for the Meissner effect, the same spin-current azimuthal speed Eq. (10) is obtained for a magnetic moment
moving radially outward a distance $2\lambda_L$ in the presence of a radial electric field Eq. (12) as for a magnetic moment orbiting at radius $2\lambda_L$ that is subject
to a time-dependent radial electric field that grows from zero  to its final value Eq. (12).

Note that the finding that the orbital angular momentum of the electron in the Cooper pair is $\hbar/2$ (Eq. (20)) was not ``built in''. Rather,
it was derived (through two equivalent routes) from the hypothesis that the size of the orbit is $2\lambda_L$, which in turn was inferred from the
existence of the Meissner effect, together with the reasonable assumption that the outgoing electron magnetic moment interacts with a positive background
of equal charge density as the charge of the superfluid ($\rho=|e|n_s$).

The magnitude of the magnetic field that will stop the spin current velocity of one of the spin orientations (the one that is parallel to the applied $\vec{B}$) satisfies, according
to Eqs. (2) and (10)
\beq
v_\sigma^0=\frac{\hbar}{4m_e\lambda_L}=-\frac{e\lambda_L}{m_ec}B_s
\eeq
hence it is given by
\beq
B_s=-\frac{\hbar c}{4e\lambda_L^2}=\frac{\Phi_0}{4\pi\lambda_L^2}
\eeq
with $\Phi_0=hc/2|e|$ the flux quantum. Eq. (29) is essentially the lower critical field of a type II superconductor, $H_{c1}$, that will drive the
system normal\cite{tinkham}, and it coincides with $B_\sigma$, Eq. (16). The flux of the ``stopping field'' $B_s$ through the area of the orbit of radius $2\lambda_L$ is precisely the
flux quantum $\Phi_0$.

\section{negative charge expulsion}
We have shown in the previous sections that expansion of the electronic orbits from a microscopic dimension to a mesoscopic radius
$2\lambda_L$ describes the Meissner effect and predicts the Spin Meissner effect\cite{sm}. This expansion also gives rise to expulsion of negative charge
from the interior of the superconductor towards the surface, as we discuss in what follows.

In the normal state, electronic orbits carry zero orbital angular momentum on average, however each electron has an intrinsic (spin) angular 
momentum $\hbar/2$. We can think of the spinning electron as a charge $e$ orbiting at speed $c$ in an orbit of radius given by the
``quantum electron radius'' $r_q\equiv \hbar/(2m_e c)$ . We have seen that as the orbit expands to radius $2\lambda_L$ the $orbital$
angular momentum acquired is also $\hbar/2$ (Eq. 20).  Thus the electron orbiting at radius $2\lambda_L$ with orbital angular momentum $\hbar/2$
can be regarded as a magnified image of the spinning electron, with `magnification factor' $2\lambda_L/r_q$. Hence 
 it is natural to conclude that the charge $e$ will be 
correspondingly $reduced$ by the 
factor $r_q/(2\lambda_L)$, so that the expelled negative charge density is
\beq
\rho_-=en_s\frac{r_q}{2\lambda_L}=en_s\frac{v_\sigma^0}{c} .
\eeq
Eq. (30) implies that the spin current can be equivalently regarded as being carried by charge densities $en_s/2$ orbiting at speed $\pm v_\sigma^0$ or by
 charge densities $\rho_-/2$ orbiting at speed $\pm c$ (the same charge density orbits in each direction in both cases in the absence of
 applied magnetic field). A similar result holds for the charge current as we discuss below.

Indeed it can be shown\cite{electrospin} that the requirement that the theory be relativistically covariant leads to the conclusion that a negative charge
density $\rho_-$ of magnitude given by  Eq. (30) exists within a London penetration depth of the surface of superconductors. 
This negative charge was expelled from the interior of the superconductor in the transition to superconductivity\cite{chargeexp},
resulting in an 
interior positive charge density
\beq
\rho_0=-\frac{2\lambda_L}{R}\rho_-
\eeq
for a cylinder of radius $R$. The electric field generated by this internal positive charge density increases linearly with $r$, the distance to the
cylinder axis, and reaches a maximum value
\beq
E_m=2\pi \rho_0 R=-\frac{\hbar c}{4e\lambda_L^2}
\eeq
within a London penetration depth of the surface. Note that  Eq. (32) is the same (in cgs units) as the
 ``stopping'' magnetic field Eq. (29) as well as the effective spin-orbit field Eq. (16).

We can also understand the result Eq. (32) from the following heuristic argument. The expelled charge density $\rho_-$ is related to $E_m$ by
\beq
\rho_-=-\frac{E_m}{4\pi\lambda_L}
\eeq
due to charge neutrality. The Meissner current in an applied magnetic field $B$ has magnitude
\beq
j=n_s |e| v_s=\frac{c}{4\pi \lambda_L}B=|\rho_-|c\frac{B}{E_m}
\eeq
Eq. (34) can be interpreted as the current created by the excess negative charge $\rho_-$ moving at speed
\beq
v_{\rho_-}=c\frac{B}{E_m}
\eeq
and suggests that superconductivity will be destroyed when $v_{\rho_-}$ reaches the speed of light. This will occur for $B=E_m$, thus the value
of the magnetic field that stops the spin current and destroys superconductivity, Eq. (29), yields the value of the electric field near the surface $E_m$ Eq. (32).

From Eqs. (3), (10) and (32) it follows that the electrostatic energy density due to the electric field $E_m$ equals the kinetic energy density of the spin current
\beq
\frac{1}{2}m_e (v_\sigma^0)^2n_s=\frac{E_m^2}{8\pi} .
\eeq
The same relation exists, as is well known\cite{tinkham}, between the kinetic energy of the Meissner current and the magnetic energy density
\beq
\frac{1}{2}m_e (v_s)^2n_s=\frac{B^2}{8\pi}.
\eeq
as can be seen from Eqs. (2) and (3).

\section{electrodynamics of charge and spin}
The foregoing considerations lead to the following four-dimensional equation in the charge sector\cite{electrodyn}
\beq
J-J_0=-\frac{c}{4\pi\lambda_L^2}(A-A_0)
\eeq
with the current four-vector given by
\beq
J \equiv (\vec{J}(\vec{r},t),ic\rho(\vec{r},t))
\eeq
with $\vec{J}$ the charge current and $\rho$ the charge density, and the vector-potential four-vector given by
\beq
A= (\vec{A}(\vec{r},t),i\phi(\vec{r},t))
\eeq
with $\vec{A}$ the magnetic vector potential and $\phi$ the electric potential, related by the Lorenz gauge condition $Div A=0$, with 
$Div\equiv(\vec{\nabla},\partial/\partial(ict))$. The quantities with subindex $0$ are
\bmath
\beq
J_0=(0,ic\rho_0)
\eeq
\beq
A_0=(0,i\phi_0(\vec{r}))
\eeq
\emath
with $\nabla^2\phi_0=-4\pi\rho_0$ and $\rho_0$ determined by Eqs. (30)-(32). The spatial part of Eq. (38) is the ordinary London equation. From 
the fourth component of Eq. (38) and Maxwell's equations it follows that there exists an electrostatic field in the
interior of superconductors that satisfies
\beq
\nabla^2(\vec{E}-\vec{E}_0)=\frac{1}{\lambda_L^2}(\vec{E}-\vec{E}_0)
\eeq
with $\vec{E}_0$ the electrostatic field generated by the uniform charge density $\rho_0$.

The charge current four-vector Eq. (39) is composed of the sum of spin current four-vectors
\beq
J=J_\uparrow+J_\downarrow
\eeq
and the spin current four-vectors satisfy\cite{electrospin}
\beq
J_\sigma-J_{\sigma 0}=-\frac{c}{8\pi\lambda_L^2} (A_\sigma-A_{\sigma 0})
\eeq
with
\bmath
\beq
J_\sigma=(\vec{J}_\sigma,ic\rho_\sigma)
\eeq
\beq
A_\sigma=(\vec{A}_\sigma,i\phi_\sigma)
\eeq
\emath
$\vec{J}_\sigma=e(n_s/2)\vec{v}_\sigma$ is the component of the current of spin $\sigma$ and $\rho_\sigma$ is the charge density with spin $\sigma$.
The spin potentials are given by\cite{electrospin}
\bmath
\beq
\vec{A}_\sigma=\lambda_L\vec{\sigma}\times\vec{E}(\vec{r},t)+\vec{A}(\vec{r},t)
\eeq
\beq
\phi_\sigma(\vec{r},t)=-\lambda_L\vec{\sigma}\cdot\vec{B}(\vec{r},t)+\phi(\vec{r},t)
\eeq
\emath 
Finally, the quantities with subindex $0$ are
\bmath
\beq
J_{\sigma 0}=(\vec{J}_{\sigma 0}(\vec{r}),ic\rho_{\sigma 0})
\eeq
\beq
\vec{J}_{\sigma 0}(\vec{r})=-\frac{c \rho_0}{2}\vec{\sigma}\times \hat{r}
\eeq
\beq
\rho_{\sigma 0}=\frac{\rho_0}{2}
\eeq
\emath
and
\bmath
\beq
A_{\sigma 0}=(\vec{A}_{\sigma 0}(\vec{r}),i\phi_{\sigma 0}(\vec{r}))
\eeq
\beq
\vec{A}_{\sigma 0}(\vec{r})=\lambda_L\vec{\sigma}\times\vec{E}_0(\vec{r})
\eeq
\beq
\phi_{\sigma 0}(\vec{r})=\phi_0(\vec{r})
\eeq
\emath
 These equations predict the existence of a spontaneous spin current flowing within a London penetration depth of the surface of the superconductor, with carrier densities
 $(n_s/2)$ and opposite spin flowing in each direction with speed Eq. (10), and a spontaneous
 electric field throughout the interior of the superconductor, of maximum value given by Eq. (32)\cite{electrospin}.

   \section{energetic considerations}
   The kinetic energy of a pair of electrons of opposite spin due to the spin current is $\epsilon_p=m_e(v_\sigma^0)^2$. When the applied magnetic field approaches $B_s$ (Eq. (29)), one of the
   members of the pair doubles its speed and the other one comes to a stop, hence the kinetic energy of the pair doubles, at which point the pair breaks up\cite{sm}. We conclude from this argument that the
   condensation energy of the pair is $m_e(v_\sigma^0)^2$ and hence that the condensation energy per electron is
   \beq
   \epsilon_c=\frac{1}{2} m_e (v_\sigma^0)^2
   \eeq
   which equals the electrostatic energy cost per electron due to the internal electric field, Eq. (36). This implies that each electron lowers its energy in entering the condensate
   by
   \beq
 \nu \equiv  2\epsilon_c=\frac{\hbar^2 q_0^2}{4m_e}
    \eeq
   with $q_0=1/2\lambda_L$. The system gives back half of this gain right away in the electrostatic energy cost Eq. (36), and the other half when the applied magnetic field destroys superconductivity.
   
   What is the physical origin of this energy lowering? Note that  $\nu$ can be written as
   \beq
   \nu=\frac{1}{2}\mu_B B_\sigma=\frac{|e|}{2m_e c^2}|\vec{S}\cdot (\vec{v}_\sigma^0\times\vec{E})|
   \eeq
   with $S=\hbar/2$ and $\vec{E}$ the radial electric field Eq. (12) at radius $r=2\lambda_L$, normal to 
   $\vec{v}_\sigma^0$. The second form of Eq. (51) is the usual spin-orbit energy including the correction for Thomas precession\cite{spinorbit}. From this
  we conclude that {\it the condensation energy of the superconductor originates in the spin-orbit energy lowering arising from the interaction of
   the spin current of the condensate (of
charge density $en_s$ and velocity Eq. (10))  with the compensating positive background charge density $|e|n_s$}.

It is also interesting to note that $\nu$ determines the fraction of the superfluid charge density ($en_s$) that is expelled (as suggested in \cite{exp}), through the
relations
\beq
\rho_- = en_s (\frac{\nu}{|e|E_m\lambda_L})=en_s (\frac{\nu}{m_e c^2})^{1/2}
\eeq
where $eE_m \lambda_L$ is the electrostatic energy difference between an electron at the center and at radius $2\lambda_L$ of a cylinder with charge density $|\rho_-|$.
(The second expression in Eq. (52) was also found in ref.\cite{chargeexp} through different arguments). $\nu$ is also related to this electrostatic energy difference through
\beq
\nu=\frac{(eE_m\lambda_L)^2}{m_e c^2}  .
\eeq

   The parameter $\nu$  represents a change in the chemical potential between normal and superconducting states\cite{exp,undr}, which according to the theory
   of hole superconductivity is related to the slope of the energy-dependent gap function $\Delta_k$ by\cite{hole7,hole8,slope}
   \beq
   \nu=\frac{1}{2}\frac{\partial}{\partial \epsilon_k} (\Delta_k)^2
   \eeq
   From Eqs. (50), (54) and (10) and using a free-electron dispersion relation $\epsilon_k=\hbar^2 k^2/2m_e$  we find that the energy-dependent gap function is given by
   \beq
   \Delta_k=\frac{\hbar^2 q_0 k}{2m_e}=\sqrt{2\nu \epsilon_k} .
   \eeq
   Both the parameter $\nu$ and the gap at the Fermi energy $\Delta_{k_F}$ can be expressed in terms of the slope of the gap function at the Fermi energy, $m$,
   that determines the tunneling asymmetry\cite{hole7}:
   \bmath \beq
   m\equiv\frac{\partial \Delta_k}{\partial \epsilon_k})_{\epsilon=\epsilon_F}=\frac{q_0}{2k_F}
   \eeq
   \beq
   \Delta_{k_F}=2m\epsilon_F
   \eeq
   \beq
   \nu=2m^2 \epsilon_F=m\Delta_{k_F}
   \eeq
   \emath
   which illustrates that a sloped gap function is a necessary condition for superconductivity\cite{hole7,slope,slope2}.     
   
   The quasiparticle energy is given within BCS theory by
   \beq
   E_k^2=(\epsilon_k-\mu)^2+\Delta_k^2=(\epsilon_k-\mu+\nu)^2+\Delta_0^2 
   \eeq
   with the minimum quasiparticle gap $\Delta_0\equiv \sqrt{2\mu \nu-\nu^2}$.  For the second equality in Eq. (57) we used Eq. (55). Eq. (57) shows that indeed
   $\nu$ is the change in the chemical potential in going from the normal to the superconducting state, as anticipated in the 
   ``correlated hopping'' model of hole superconductivity\cite{hole8,hole7,exp}. 

             \begin{figure}
\resizebox{8.5cm}{!}{\includegraphics[width=7cm]{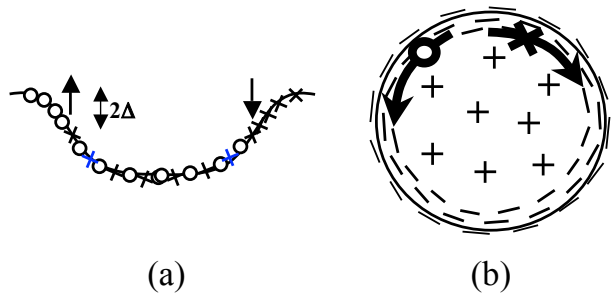}}
\caption{ Schematic depiction of spin-split bands (a) and associated real-space picture (b). (a) The spin up electrons (depicted as circles) occupy predominantly
negative k-s and the spin down electrons (depicted as crosses) positive k's. The difference in energy between the highest occupied k-state for positive and negative k for
a given spin orientation is twice the energy gap. (b) The excess negative charge near the surface of the superconductor 
flows predominantly clockwise or counterclockwise depending on its spin orientation. }
\label{atom6}
\end{figure}

Furthermore note that a spin current with speed Eq. (10) can be represented by the energy-wavevector relation\cite{hu}
   \beq
 \epsilon_{k\sigma}=\epsilon_{-k,-\sigma}=\frac{\hbar^2}{2m_e}(\vec{k}-\sigma\frac{\vec{q}_0}{2})^2
   \eeq
   with $\sigma=\pm 1$, where $\vec{q}_0$ is a vector in the direction of the spin current flow of magnitude $q_0=1/2\lambda_L$
   (since $\hbar^{-1}(\partial \epsilon_{k\uparrow}/\partial k-\partial \epsilon_{k\downarrow}/\partial k)/2=\hbar q_0/(2m_e)=v_\sigma^0$). Eqs. (58) and (55) then imply that
   \beq
   \Delta_k=\frac{\epsilon_{k\uparrow}-\epsilon_{-k\uparrow}}{2}=\frac{\epsilon_{k\uparrow}-\epsilon_{k\downarrow}}{2}
   \eeq
   for $\vec{k}\parallel \vec{q}_0$. In other words, {\it the superconducting energy gap is due to `spin splitting'} \cite{spinsplit}.
   Note also  that the `depairing' speed which will cause the spin current to stop and the pairs to break is given by Eq. (10), which can be written as
   \beq
   v_\sigma^0=\frac{\Delta_{k_F}}{\hbar k_F}
   \eeq
   in terms of the gap Eq. (55) at $k=k_F$. Eq. (60) for the critical speed is identical to what is obtained in  conventional BCS theory\cite{tinkham}, 
   even though the
   energy gap expression Eq. (55) was obtained through an entirely independent argument.
   Fig. 1 shows schematically the spin-split bands in momentum space and the associated charge configuration and spin current in real space.

       \begin{figure}
\resizebox{8.5cm}{!}{\includegraphics[width=7cm]{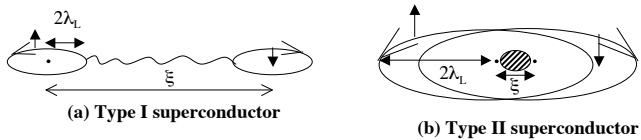}}
\caption{ Schematic depiction of a  Cooper pair in  a type I (a) and type II (b) superconductor.  The vertical arrows denote the direction of the electron magnetic moment, and the
horizontal arrows the orbiting direction. The radius of the orbit of each electron is $2\lambda_L$, and the distance between the
centers of the orbits  is $\xi$. 
In type II materials ((b)) with $\xi<2\lambda_L$  a normal vortex core of diameter $\xi$ can be enclosed by both orbits of the same Cooper pair.
}
\label{atom6}
\end{figure}
   
Finally, note that the condensation energy per electron Eq. (49) can be written, using Eq. (55) and the free-electron dispersion relation, as
   \beq
   \epsilon_c=\frac{\Delta_{k_F}^2}{4\epsilon_F}
   \eeq
   Eq. (61) is consistent with the BCS expression for the condensation energy per unit volume\cite{tinkham}
   \beq
   \delta U=\frac{1}{2}N(0)\Delta^2
   \eeq
   with $N(0)$ the density of states per spin at the Fermi energy, if we take $N(0)=n_s/2\epsilon_F$ appropriate to a two-dimensional free-electron system.
   
   Van der Marel\cite{marel} and Khomskii\cite{khomskii} have pointed out that quite generally in conventional BCS theory a shift in the chemical potential
   is predicted upon entering the superconducting state, of magnitude $\Delta^2/4\epsilon_F$. In our case the shift in the chemical potential, $\nu$, is twice
   as large (Eqs. (50) and (61)), and it is directly related to the slope of the gap function and to the existence of a spin current and of negative charge expulsion.

  \section{type I versus type II materials and charge inhomogeneity}
  The above energetic considerations apply close to the crossover between type I and type II behavior, where $H_c\sim H_{c1}$ and $\lambda_L\sim\xi$, where
  $\xi$ is the coherence length which is also the average distance between members of a Cooper pair\cite{tinkham}. In extreme type I materials,
  pairs are broken well before the charge speed reaches Eq. (10), $q_0\sim 1/\xi<<1/2\lambda_L$ in Eq. (55) and
  $E_m=H_c$ rather than Eq. (32). We can understand the crossover between type I and type II behavior geometrically, as shown in Fig. 2.
  $\xi$ is the distance between the centers of the $2\lambda_L$ orbits of up and down spin electrons in a Cooper pair. A vortex core has diameter $\xi$, and this normal region can be enclosed by the orbits of both members of the same Cooper pair if and
  only if $\xi<2\lambda_L$.

     \begin{figure}
\resizebox{8.0cm}{!}{\includegraphics[width=7cm]{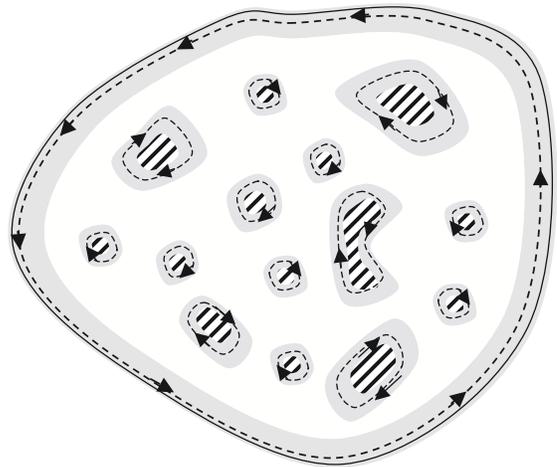}}
\caption{Schematic depiction of a superconductor with strong disorder in the absence of applied magnetic field. Defects, grain boundaries, vacancies, etc. will result
in patches of normal regions (hatched areas) surrounded by spin currents (dashed lines, with arrows pointing in the direction of flow of electrons with
magnetic moment pointing $out$ of the paper) and excess negative charge density (gray areas). 
The figure also shows the excess negative charge and spin current near the surface. 
If the system is cooled in the presence of a magnetic field, magnetic flux will be trapped in the hatched regions and a charge current will flow around
those regions together
with the depicted  spin currents. The smallest normal regions have diameter of a coherence length.}
\label{atom5}
\end{figure}
  
  Why does the vortex core have to be enclosed by the orbits of both members of the same Cooper pair? The phase change for each electron in going around
  a loop enclosing a vortex is
  \beq
  \hbar \Delta \theta=\oint  m_e\vec{v} \cdot d\vec{l}+\frac{e}{c}\phi_B
  \eeq
  where $\phi_B$ is the enclosed magnetic flux. For each member of the Cooper pair this phase change is $\pi$, corresponding to its angular momentum
  $\hbar/2$, and thus $\phi_B=(h/2)c/e=\phi_0$ (assuming the integration loop is through a path where $\vec{v}=0$), thus
  providing a new rationale for the factor of 2 in the flux quantum $\phi_0$. If only one member of the Cooper pair
were to enclose the vortex the wave function for the pair would not be single-valued.

 The superconductor expels negative charge towards the surface and towards any interior normal regions, thus in the flux phase there will be excess negative charge
  in and around the vortex cores.    
  Furthermore our model has a large sensitivity to disorder arising from the slope of the gap function, as discussed in \cite{local}. In the presence of local potential variations
  due to impurities, vacancies, etc, the gap can be sharply reduced or vanish altogether giving rise to normal regions. There will be excess negative
  charge in and around those normal regions expelled from the superconducting regions, and a spin current will circulate around the interior normal regions
  as shown schematically in Fig. 2. An applied magnetic field will concentrate in these normal or weakly superconducting regions and the spin current around them
  will acquire a charge current component.

Finally, the model has a strong tendency to phase separation in the regime where the carrier (hole) concentration is small\cite{ps}.
This arises from the bandwidth dependence on the hole concentration\cite{bandwidth}: the bandwidth increases with increasing hole concentration,
favoring segregation into hole-rich and hole-poor regions. This tendency is also enhanced in the presence of disorder.

  \section{summary and discussion}
  
 The theory discussed here proposes that superconductivity arises from the fundamental charge asymmetry of matter\cite{chargeasymm}. It was motivated by the 
 discovery of high $T_c$ cuprates\cite{hole4}
 but it was clear from the outset that if valid it would  apply to all  superconducting materials\cite{first}. As the theory progressed it became
 increasingly apparent that it led to a radical departure from the conventional BCS theory\cite{hole5,undr}. Later it became clear that even London electrodynamics had to be
 modified\cite{electrodyn}. Finally it became clear that   spin-orbit coupling plays a key role\cite{sm,electrospin}. The fact that at the end of this road one is led, unexpectedly, to a 
far more compelling  (in our view) explanation\cite{missing} of the 
 most fundamental  property of superconductors, the Meissner effect, than the conventional theory proposes, is in our view a strong argument for its validity.
Ultimately of course confirmation or refutation of the theory 
discussed here will come from experiment.
 
 As the theory advanced we have gained increased understanding of what we knew we didn't know, however we have also become aware of issues that previously had been  
 'unknown unknowns'\cite{rums}. Perhaps the
 most important of these issues, that remains  to be understood, is the proper inclusion of the key role of the Dirac sea\cite{dirac}.

\acknowledgements
I am grateful to F. Marsiglio for collaboration in the initial development as well as later stages of this theory.
 
\end{document}